\documentclass[12pt]{article}
\usepackage{epsfig,graphicx}
\usepackage{fpcp04}

\def\beq{\begin{equation}}
\def\eeq#1{\label{#1}\end{equation}}
\def\eeqn{\end{equation}}

\def\beqa{\begin{eqnarray}}
\def\eeqa#1{\label{#1}\end{eqnarray}}
\def\eeqan{\end{eqnarray}}

\let\bar=\overbar

\def\Dslash{\not{\hbox{\kern-4pt $D$}}}
\def\dslash{\not{\hbox{\kern-2pt $\del$}}}

\def\msb{{\bar{\ssstyle M \kern -1pt S}}}

\def\BB0bar{B^0 {\overline B}^0}
\def\BB0dbar{B_d^0 {\overline B}_d^0}
\def\BB0sbar{B_s^0 {\overline B}_s^0}

\RequirePackage{xspace}

\usepackage{relsize}
\def\babar{\mbox{\slshape B\kern-0.1em{\smaller A}\kern-0.1em
    B\kern-0.1em{\smaller A\kern-0.2em R}}}

\def\Kbar  {\kern 0.2em\overline{\kern -0.2em K}{}\xspace}

\def\Kz    {\ensuremath{K^0}\xspace}
\def\Kzb   {\ensuremath{\Kbar^0}\xspace}
\def\KzKzb {\ensuremath{\Kz \kern -0.16em \Kzb}\xspace}
\def\Kp    {\ensuremath{K^+}\xspace}
\def\Km    {\ensuremath{K^-}\xspace}

\def\KpKm  {\ensuremath{\Kp \kern -0.16em \Km}\xspace}

\def\Dbar    {\kern 0.2em\overline{\kern -0.2em D}{}\xspace}

\def\Dz      {\ensuremath{D^0}\xspace}
\def\Dzb     {\ensuremath{\Dbar^0}\xspace}
\def\DzDzb   {\ensuremath{\Dz {\kern -0.16em \Dzb}}\xspace}
\def\Dp      {\ensuremath{D^+}\xspace}
\def\Dm      {\ensuremath{D^-}\xspace}

\def\DpDm    {\ensuremath{\Dp {\kern -0.16em \Dm}}\xspace}

\def\Bbar    {\kern 0.18em\overline{\kern -0.18em B}{}\xspace}

\def\BB      {\ensuremath{B\Bbar}\xspace} 
\def\Bz      {\ensuremath{B^0}\xspace}
\def\Bzb     {\ensuremath{\Bbar^0}\xspace}
\def\BzBzb   {\ensuremath{\Bz {\kern -0.16em \Bzb}}\xspace}
\def\Bu      {\ensuremath{B^+}\xspace}
\def\Bub     {\ensuremath{B^-}\xspace}

\def\BpBm    {\ensuremath{\Bu {\kern -0.16em \Bub}}\xspace}

\mathchardef\Upsilon="7107
\def\Y#1S{\ensuremath{\Upsilon{(#1S)}}\xspace}

\mathchardef\Deltares="7101
\mathchardef\Xi="7104
\mathchardef\Lambda="7103
\mathchardef\Sigma="7106
\mathchardef\Omega="710A

\def\Deltabar{\kern 0.25em\overline{\kern -0.25em \Deltares}{}\xspace}
\def\Lbar{\kern 0.2em\overline{\kern -0.2em\Lambda\kern 0.05em}\kern-0.05em{}\xspace}
\def\Sigbar{\kern 0.2em\overline{\kern -0.2em \Sigma}{}\xspace}
\def\Xibar{\kern 0.2em\overline{\kern -0.2em \Xi}{}\xspace}
\def\Obar{\kern 0.2em\overline{\kern -0.2em \Omega}{}\xspace}
\def\Nbar{\kern 0.2em\overline{\kern -0.2em N}{}\xspace}
\def\Xb{\kern 0.2em\overline{\kern -0.2em X}{}\xspace}

\newcommand{\tev}{\ensuremath{\mathrm{\,Te\kern -0.1em V}}\xspace}
\newcommand{\gev}{\ensuremath{\mathrm{\,Ge\kern -0.1em V}}\xspace}
\newcommand{\mev}{\ensuremath{\mathrm{\,Me\kern -0.1em V}}\xspace}
\newcommand{\kev}{\ensuremath{\mathrm{\,ke\kern -0.1em V}}\xspace}
\newcommand{\ev}{\ensuremath{\mathrm{\,e\kern -0.1em V}}\xspace}
\newcommand{\gevc}{\ensuremath{{\mathrm{\,Ge\kern -0.1em V\!/}c}}\xspace}
\newcommand{\mevc}{\ensuremath{{\mathrm{\,Me\kern -0.1em V\!/}c}}\xspace}
\newcommand{\gevcc}{\ensuremath{{\mathrm{\,Ge\kern -0.1em V\!/}c^2}}\xspace}
\newcommand{\mevcc}{\ensuremath{{\mathrm{\,Me\kern -0.1em V\!/}c^2}}\xspace}

\def\mus  {\ensuremath{\rm \,\mus}\xspace}

\def\mus        {\ensuremath{\,\mu{\rm s}}\xspace}    

\def\to                 {\ensuremath{\rightarrow}\xspace}

\def\pep2{PEP-II}

\def\gsim{{~\raise.15em\hbox{$>$}\kern-.85em
          \lower.35em\hbox{$\sim$}~}\xspace}
\def\lsim{{~\raise.15em\hbox{$<$}\kern-.85em
          \lower.35em\hbox{$\sim$}~}\xspace}

\def\Vud  {\ensuremath{|V_{ud}|}\xspace}
\def\Vcd  {\ensuremath{|V_{cd}|}\xspace}
\def\Vtd  {\ensuremath{|V_{td}|}\xspace}
\def\Vus  {\ensuremath{|V_{us}|}\xspace}
\def\Vcs  {\ensuremath{|V_{cs}|}\xspace}
\def\Vts  {\ensuremath{|V_{ts}|}\xspace}
\def\Vub  {\ensuremath{|V_{ub}|}\xspace}
\def\Vcb  {\ensuremath{|V_{cb}|}\xspace}
\def\Vtb  {\ensuremath{|V_{tb}|}\xspace}

\xspace

\def\jetset74   {\mbox{\tt Jetset \hspace{-0.5em}7.\hspace{-0.2em}4}\xspace}

\def\to{\rightarrow}

\let\<=\langle
\let\>=\rangle

\def\beq{\begin{equation}}
\def\eeq{\end{equation}}
\def\ba{\begin{array}}
\def\bea{\begin{eqnarray}}
\def\ea{\end{array}}
\def\eea{\end{eqnarray}}
\def\bit{\begin{itemize}}
\def\eit{\end{itemize}}

\def\comment#1{ \hbox{[{\it Comment suppressed here.}\/]} }
\def\hide#1{}
      
\def\O{ {\cal O} }

\newcommand{\nn}{\nonumber }

\def\Vud{0.9744(5)(3)}
\def\Vus{~~~0.225(2)(1)~~~}
\def\Vub{3.5(5)(5)\!\times\! 10^{-3}}
\def\Vcd{~~~0.24(3)(2)~~~}
\def\Vcs{~~~~0.97(10)(2)~~~~}
\def\Vcb{3.9(1)(3)\!\times\! 10^{-2}}
\def\Vtd{8.1(2.7)\!\times\! 10^{-3}}
\def\Vts{3.8(4)(3)\!\times\! 10^{-2}}
\def\Vtb{0.9992(0)(1)}
\def\Wlambda{\Vus}
\def\WA{0.77(2)(7)}
\def\WR{0.40(6)(6)}
\def\Wrho{0.16(28)}
\def\Weta{0.36(11)}

\begin{document}

\Title{Full CKM matrix with lattice QCD}
\bigskip

%
\label{MOkamotoStart}

%
\author{ Masataka Okamoto\index{Okamoto, M.}\\
(for the Fermilab Lattice, MILC, and HPQCD Collaborations) }

%
\address{
Fermi National Accelerator Laboratory, 
Batavia, Illinois 60510, USA \\
e-mail: okamoto@fnal.gov
}

\makeauthor\abstracts{
We show that it is now possible to fully determine
the CKM matrix, for the first time, using lattice QCD.
$|V_{cd}|$, $|V_{cs}|$, $|V_{ub}|$, $|V_{cb}|$ and $|V_{us}|$
are, respectively, directly determined
with our lattice results for form factors of
semileptonic $D\to \pi l\nu$,
             $D\to K l\nu$,
             $B\to \pi l\nu$,
             $B\to D l\nu$
         and $K\to \pi l\nu$ decays.
The error from the quenched approximation is removed by
using the MILC unquenched lattice gauge configurations,
where the effect of $u,d$ and $s$ quarks is included.
The error from the ``chiral'' extrapolation ($m_l\to m_{ud}$) 
is greatly reduced by using improved staggered quarks.
The accuracy is comparable to 
that of the Particle Data Group averages. 
In addition,
$|V_{ud}|$, $|V_{tb}|$, $|V_{ts}|$ and $|V_{td}|$ are determined
by using unitarity of the CKM matrix and the experimental result
for $\sin{(2\beta)}$.
In this way, we obtain all 9 CKM matrix elements, where 
the only theoretical input is lattice QCD. 
We also obtain all the Wolfenstein parameters,
for the first time, using lattice QCD.
}

\section{Introduction}

The Cabibbo-Kobayashi-Maskawa (CKM) matrix,
which relates the mass eigenstates and the weak eigenstates
in the Standard Model electroweak theory, is a set of parameters.
To determine each CKM matrix element, one requires both
theoretical and experimental inputs.
On the theoretical side, one needs to know relevant hadronic
amplitudes, which often contain nonperturbative QCD effects.
A major role of lattice QCD is to calculate such hadronic
amplitudes reliably and accurately, from first principles.
One can then extract the CKM matrix elements by combining 
lattice QCD as the theoretical input
with the experimental input such as decay rates.
In this paper, we show that it is now possible to {\it fully} determine
the CKM matrix, for the first time, using lattice QCD.
The result for the full CKM matrix with lattice QCD is:
        \bea V_{\rm CKM} ~=~
        \left(
        \begin{array}{ccc}
        {{|V_{ud}|}}   & {|V_{us}|}   & {|V_{ub}|} \\
            \Vud       &    \Vus      &    \Vub    \\
        { |V_{cd}| }   & {|V_{cs}|}   & {|V_{cb}|} \\
            \Vcd       &    \Vcs      &    \Vcb    \\
        { |V_{td}| }   & {|V_{ts}|}   & {|V_{tb}|} \\
            \Vtd       &    \Vts      &    \Vtb    \\
        \end{array}
        \right) \label{ckm}
        \eea
where the first errors are from lattice calculations
and the second are experimental, except the one for 
$|V_{td}|$ which is a combined lattice and experimental error.
The results for the Wolfenstein parameters with lattice QCD are:
\bea
\lambda = \!\!\!\!\Wlambda\!\!\!\!\!\!,~~~~
A = \WA ,~~~~
\rho = \Wrho ,~~~~
\eta = \Weta.
\label{Wolf}
\eea

To directly determine 5 CKM matrix elements 
($|V_{cd}|$, $|V_{cs}|$, $|V_{ub}|$, $|V_{cb}|$ and $|V_{us}|$),
we use 5 semileptonic decays 
($D\to \pi l\nu$,
 $D\to K l\nu$,
 $B\to \pi l\nu$,
 $B\to D l\nu$ and
 $K\to \pi l\nu$),
for which the techniques for lattice calculations are well
established, and thus reliable calculations are possible.
The accuracy of previous lattice calculations
was limited
by two large systematic uncertainties ---
the error from the ``quenched'' approximation (neglect of
virtual quark loop effects)
and the error from the ``chiral'' 
extrapolation in light quark mass ($m_l\to m_{ud}$).
Both led to effects of around 10--20\%.
Our present work~\cite{Aubin:2004ej,Okamoto:2004xg} 
successfully reduces these two dominant uncertainties.
The error from the quenched approximation is removed by
using the MILC unquenched gauge configurations~\cite{milc},
where the effect of $u,d$ and $s$ quarks is included ($n_f=2+1$).
The error from the chiral extrapolation is greatly 
reduced by using improved staggered quarks. 
With this improved approach, the accuracy of the 5 CKM matrix elements
is comparable to that of the Particle Data Group~\cite{Eidelman:wy}. 
The results for $|V_{ub}|$, $|V_{cb}|$ and $|V_{us}|$ are preliminary.

Since we determine all 3 elements in the charm row of the CKM matrix,
$|V_{cq}|~(q=d,s,b)$, we can check   
a unitarity condition on the CKM matrix using only 
our results from lattice QCD. 
Adding in $|V_{us}|$ and $|V_{ub}|$, we then use CKM unitarity
to determine the other
4 CKM matrix elements ($|V_{ud}|$, $|V_{tb}|$, $|V_{ts}|$ and $|V_{td}|$).
In this way, we obtain all 9 CKM matrix elements 
and all the Wolfenstein parameters,
where 
the only theoretical input is lattice QCD.\footnote{We assume here that 
the Standard Model is correct. We plan to test 
the Standard Model in future work.
}

This paper is organized as follows.
In Sec.~\ref{sec:SLdecay} we present our results
for the 5 semileptonic decays and the 5 CKM matrix elements.
In Sec.~\ref{sec:unitary} we first check unitarity
of the charm row of the CKM matrix,
and then give the results for the other 4 CKM matrix elements from  
unitarity, as well as the Wolfenstein parameters.
In Sec.~\ref{sec:summary} we give a summary and discuss future plans.
The results for the $D$ and $B$ decays have been presented in 
Refs.~\cite{Aubin:2004ej,Okamoto:2004xg}.
This work is a part of our ongoing project 
of heavy quark physics with lattice QCD; see also
\cite{Simone:2004fr,diPierro:2003iw}.

\section{5 CKM matrix elements from 5 semileptonic decays}
\label{sec:SLdecay}

\subsection{$D\to\pi(K)l\nu$, $|V_{cd(s)}|$ and $B\to\pi l\nu$, $|V_{ub}|$}
The differential decay rate $d\Gamma/dq^2$ for 
the heavy-to-light semileptonic decay $H\to Pl\nu$
is proportional to $|V_{ij}|^2 |f_+(q^2)|^2$,
where $f_+$ is a form factor of the relevant hadronic amplitude
defined through
\bea
\< P | V^\mu | H \>
 &=& 
f_+(q^2) (p_H+p_P-\Delta)^\mu + f_0(q^2) \Delta^\mu \label{eq:HLff}\\
&=& 
\sqrt{2m_H} \, \left[v^\mu \, f_\parallel(E) +
p^\mu_\perp \, f_\perp(E) \right]. \nn
\eea
Here $q = p_H - p_P$, $\Delta^\mu=(m_H^2-m_P^2)\, q^\mu / q^2$,
$v=p_H/m_H$, $p_\perp=p_P-Ev$ and $E=E_P$.
To determine the CKM matrix element $|V_{ij}|$ with the
experimental rate
$\int^{q^2_{\rm max}}_{q^2_{\rm min}}dq^2\ (d\Gamma/dq^2)$,
we calculate $f_{+,0}$ as a function of $q^2$.
Below we briefly describe our analysis 
procedure~\cite{Aubin:2004ej,Okamoto:2004xg}.

We first extract the
form factors $f_\parallel$ and $f_\perp$,
and carry out the chiral extrapolation in $m_l$
for them at fixed final-state energy $E$.
To this end, we interpolate and
extrapolate the results for $f_\parallel$ and $f_\perp$ to common values
of $E$ using 
the parametrization of Becirevic and Kaidalov (BK)
\cite{Becirevic:1999kt}. 
We perform the chiral extrapolation ($m_l\to m_{ud}$) at each $E$
using the NLO correction
in staggered chiral perturbation theory~\cite{Aubin:2004xd}.
We try various fit forms for the extrapolation, 
and the differences between the fits, 2--4\%, are taken as 
associated systematic errors. 

\begin{figure}[tb]
\begin{center}
\centerline{
\epsfig{file=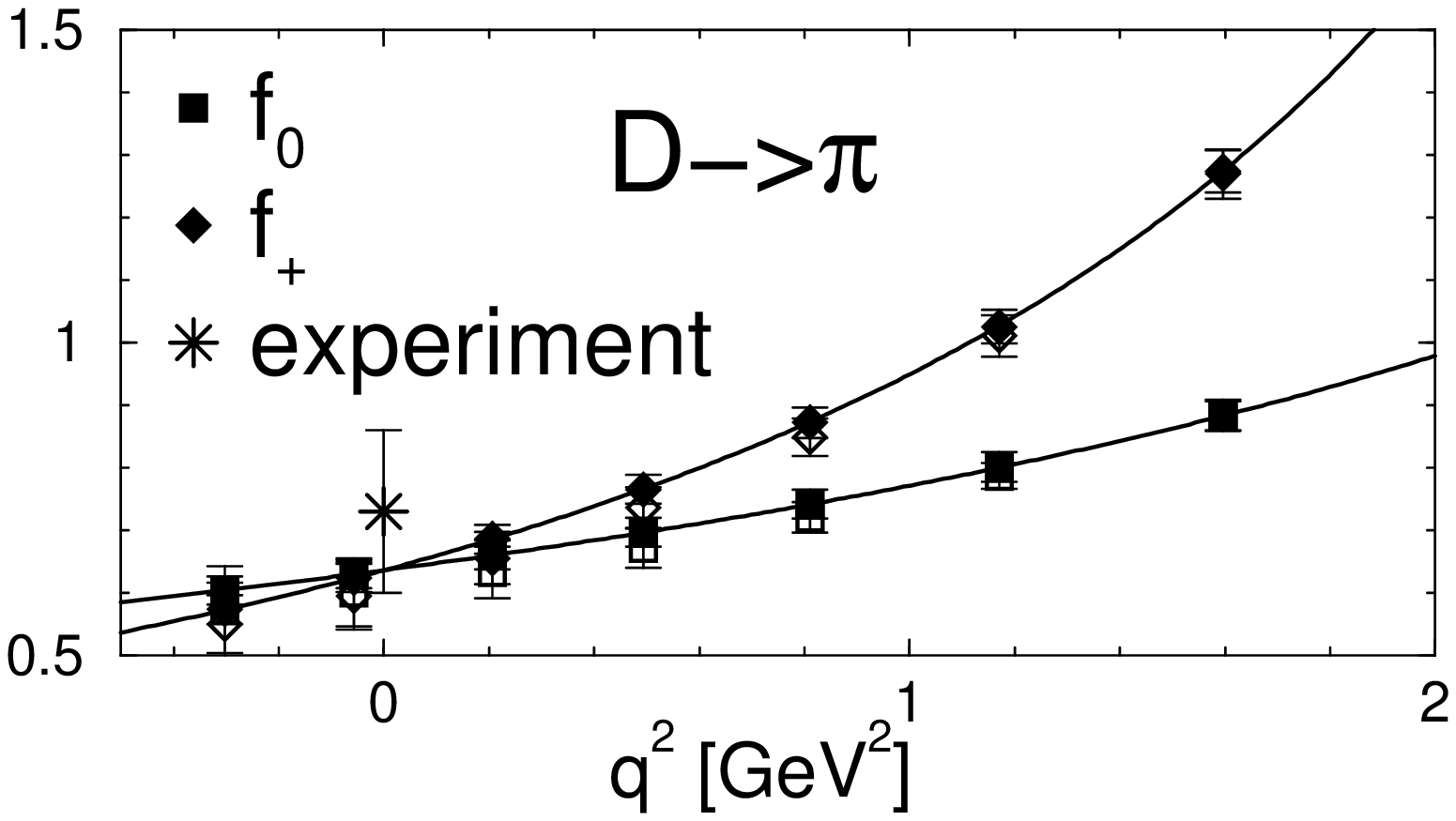,height=50mm}
\epsfig{file=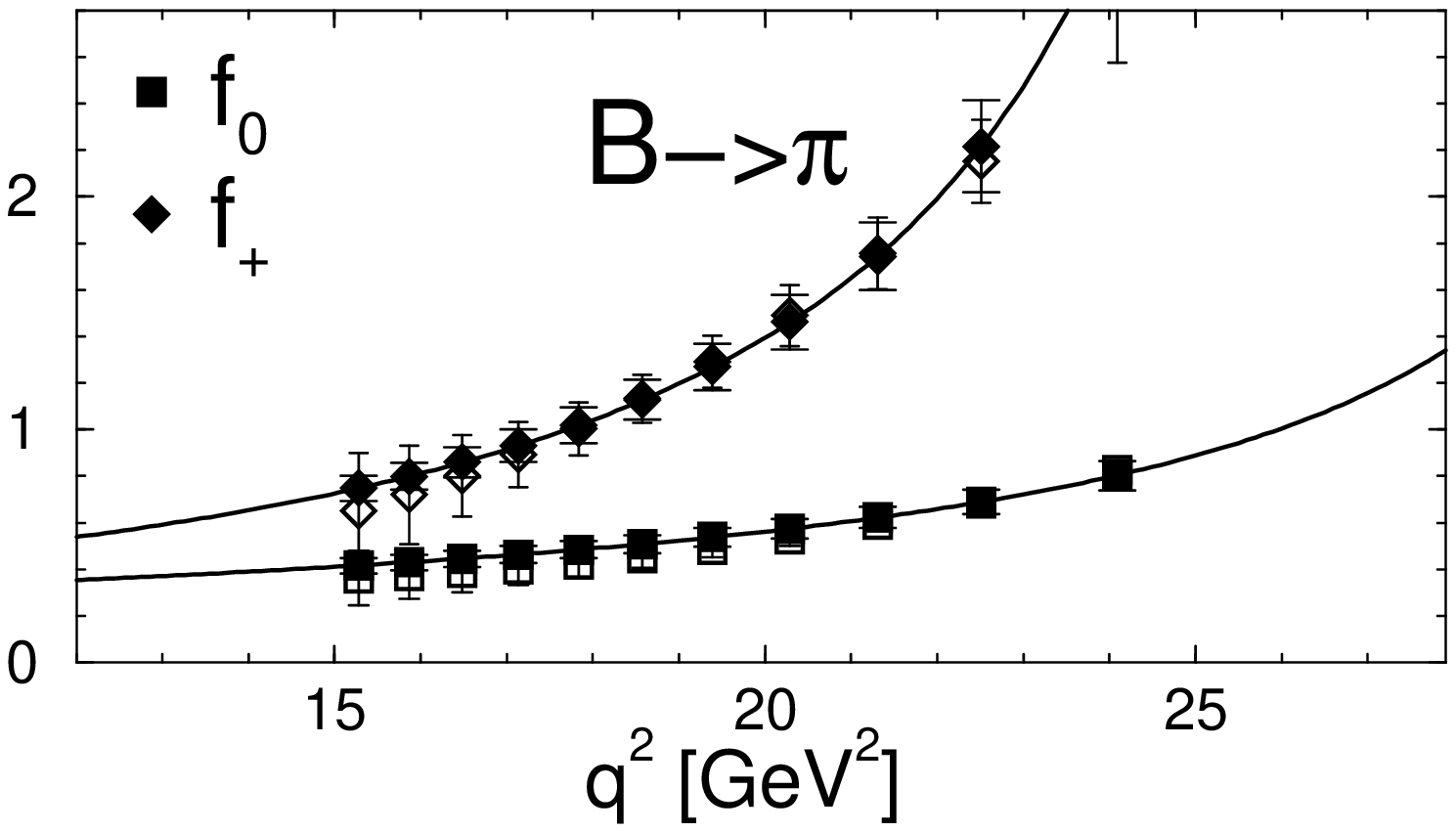,height=50mm}}
\vspace{-.7cm}
\caption{Form factors from 
BK-based (filled symbols and curves) and non-BK-based (open symbols) analyses
for $D\to\pi l\nu$ (left panel) and $B\to\pi l\nu$ (right) decays.}
\label{fig:B2pi}
\vspace{-1cm}
\end{center}
\end{figure}

We then convert 
the results for $f_{\perp}$ and $f_{\parallel}$ at $m_l=m_{ud}$,
to $f_+$ and $f_0$.
To extend
$f_+$ and $f_0$ to functions of $q^2$,
we again make a fit using the BK parameterization \cite{Becirevic:1999kt},
\bea\label{eq:BK}
f_+(q^2) = \frac{f_+}{(1-\tilde{q}^2)(1-\alpha\tilde{q}^2)},~~~
f_0(q^2) = \frac{f_+}{1-\tilde{q}^2/\beta},
\eea
where $\tilde{q}^2=q^2/m_{H^{*}}^2$.
We obtain~\cite{Aubin:2004ej,Okamoto:2004xg}
\bea
f_+^{D\to \pi}=0.64(3),~&
\alpha^{D\to  \pi}=0.44(4),~& \beta^{D\to \pi}=1.41(6), \\
f_+^{D\to  K}=0.73(3),~  &
\alpha^{D\to  K}=0.50(4),~  & \beta^{D\to  K}=1.31(7),
\eea
for the $D$ decay, and 
\bea
f_+^{B\to\pi}=0.23(2),~ &\alpha^{B\to\pi}=0.63(5),~ & \beta^{B\to\pi}=1.18(5),
\eea
for the $B$ decays, where the errors are statistical only. 
To estimate the error from the BK parameterization, {\it i.e.,}
the error for $q^2$ dependence,
we also make an alternative analysis, where we perform a 2-dimensional 
polynomial fit in $\left(m_l,E(q^2)\right).$ 
A comparison between two analyses is shown in Fig.~\ref{fig:B2pi}.
The results for $D$ decays agree well with recent experimental 
results \cite{unknown:2004nn}.

We now determine the CKM matrix elements 
by integrating $|f_+(q^2)|^2$ over $q^2$ and 
using experimental decay rates~\cite{Eidelman:wy,Athar:2003yg,Belle:B2pi}.
For $|V_{ub}|$ we use a combined average of 
the decay rate for $q^2 \ge 16$ GeV$^2$
in Refs.~\cite{Athar:2003yg} and \cite{Belle:B2pi}.
We obtain
\bea
~~~|V_{cd}| = 0.239(10)(24)(20) \,\,\,\,\, ,\,\,\,\,\,
   |V_{cs}| = 0.969(39)(94)(24)
\label{eq:VcdVcs}
\eea
from the $D$ decay, and 
\bea
~~~|V_{ub}|\times 10^{3} = 3.48(29)(38)(47) 
\label{eq:Vub}
\eea
from the $B$ decays, where the first errors are statistical,
the second systematic, and the third are experimental errors from
the decay rates.
The systematic errors 
are dominated by the finite lattice spacing effects,
{\it i.e.,} the lattice discretization effects; 
see Table~\ref{tab:error}.
The results for the CKM matrix elements agree with the 
Particle Data Group averages~\cite{Eidelman:wy} with a comparable accuracy.
A similar calculation by the HPQCD collaboration is also
underway~\cite{Shigemitsu:2004ft}.

\begin{table}[b]
\begin{center}
\begin{tabular}{l|rrrr}
\hline
semileptonic decay  &$D\to \pi(K)l\nu$ & $B\to \pi l\nu$ & $B\to D l\nu$ & 
$K\to \pi l\nu $\\
CKM matrix element  &$|V_{cd(s)}|$  &$|V_{ub}|$   &$|V_{cb}|$ & $|V_{us}|$\\
\hline
$q^2$ dependence & 2\% & 4\% &  & $<$1\%\\
$m_l\!\to\! m_{ud}$ extrapolation& 3\%(2\%) &4\% & 1\% & 1\%\\
operator matching & $<$1\% & 1\% &  1\%  & $<$1\%\\
{discretization effects} & { 9\%}& { 9\%} &  $<$1\%  & \\
\hline
total systematic error   & { 10\%} & { 11\%} &  { 2\%}  & {1\%}\\
\hline
\hline
\vspace{-.4cm}\\
error in PDG average
& 5\%(1\%)       & 13\%       &  4\%    & 1\%\\
\hline
\end{tabular}
\caption{Systematic errors in lattice calculations. 
For comparison, the error for each CKM matrix element by the
Particle Data Group~\cite{Eidelman:wy} is shown in the last row.}
\label{tab:error}
\end{center}
\end{table}

\subsection{$B\to D l\nu$, $|V_{cb}|$ and $K\to \pi l\nu$, $|V_{us}|$}

The form factors of $B\to D l\nu$ and $K\to \pi l\nu$ decays
can be calculated more accurately than those of 
heavy-to-light ($H\to P l\nu$) decays, due to the symmetry
between the initial and final states.
The differential decay rate of $B\rightarrow D l\nu$ is
proportional to the square of $|V_{cb}| {\cal{F}}(w)$, 
where ${\cal{F}}(w)$ is 
a linear combination of $h_+(w)$ and $h_-(w)$ defined through
\begin{eqnarray}\label{eq:definition_of_the_form_factors}
\langle D| V^{\mu}|{{B}}\rangle ~=~ \sqrt{m_B m_D} \times 
[h_+(w) (v+v')^{\mu} + h_-(w) (v-v')^{\mu}],
\end{eqnarray}
with $v=p_B/m_B$, $v'=p_D/m_D$ and
$w=v\cdot v'$.
To extract $|V_{cb}|$,
we calculate the form factor at $w=1$, ${\cal{F}}(1)$,
by employing the double ratio method \cite{Hashimoto:1999yp}.
The light quark mass dependence of ${\cal{F}}(1)$ is 
mild, and by
extrapolating the result linearly to $m_l\to 0$ 
we obtain~\cite{Okamoto:2004xg}
\bea
{\cal{F}}^{B\rightarrow D}(1)=1.074(18)(16),
\label{b2dFF}
\eea
where the first error is statistical, and the second is systematic, as
summarized in Table~\ref{tab:error}.
Using our result Eq.~(\ref{b2dFF}) and an average of experimental results for 
$|V_{cb}{|\cal{F}}(1)$~\cite{HFAG}, we obtain 
\bea
   |V_{cb}|\times 10^{2} = 3.91(07)(06)(34),
\label{eq:Vcb}
\eea
where the errors from the lattice calculation (first two)
are much smaller than the experimental one (third).

\begin{figure}[tb]
\begin{center}
\epsfig{file=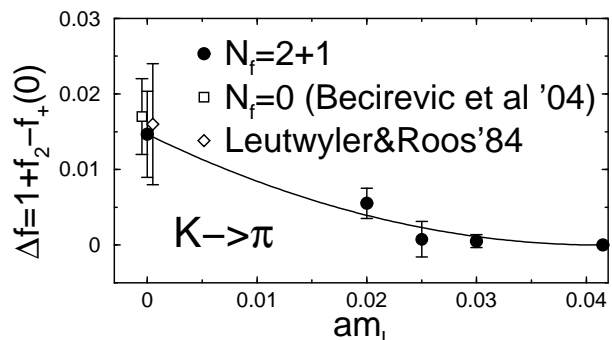,height=50mm}
\vspace{-.7cm}
\caption{$m_l$-dependence of $\Delta f$ for $K\to\pi l\nu$ decay,
together with results from Refs.~\cite{Becirevic:2004ya,Leutwyler:1984je}.}
\label{fig:K2pi}
\vspace{-.8cm}
\end{center}
\end{figure}

Finally we study the $K\to\pi l\nu$ decay to determine $|V_{us}|$.
The expression for the $K\to\pi$ decay amplitude is given in an
analogous way to Eq.~(\ref{eq:HLff}).
We calculate the $K^0\to\pi^-$ form factor at $q^2=0$, $f_+(0)=f_0(0)$,
by employing the following three steps method, as in 
Ref.~\cite{Becirevic:2004ya}:
\begin{enumerate}
\item Extract the $f_0$ form factor at $q^2=q^2_{\rm max}$
by applying the double ratio method similar to the $B\to D$ 
case~\cite{Hashimoto:1999yp}.
\item Extrapolate $f_0$ at $q^2=q^2_{\rm max}$ to $q^2=0$
by using the experimental result~\cite{Alexopoulos:2004sw}
for the slope parameter
$\lambda_0$ defined by $f_0(q^2)\!=\!f_0(0)/(1-\lambda_0 q^2)$.\footnote{
An unquenched lattice calculation of 
$\lambda_0$ should be done in the future.}
This gives $f_0(0)\!=\!f_+(0)$ for each $m_l$.
\item Perform the $m_l\!\to\! m_{ud}$ extrapolation for $f_+(0)$.
To this end, we subtract the leading logarithmic correction $f_2$
in chiral perturbation theory,
{\it i.e.,} define $\Delta f\equiv 1+f_2-f_+(0)$.
We make a fit to $\Delta f$ adopting an ansatz,
$\Delta f = (A +B m_l)(m_s-m_l)^2$, where $A,B$ are fit parameters.
The $(m_s-m_l)^2$ dependence is expected due to the Ademollo-Gatto theorem.
\end{enumerate}
As an exploratory study of ${f}_{+}^{K\to\pi}$, 
we use an improved staggered action
for $u,s$ quarks and an improved Wilson action for the $d$ quark.
The $m_l$-dependence of $\Delta f$ and the extrapolated result
are shown in Fig.~\ref{fig:K2pi}, together with 
a recent quenched lattice result~\cite{Becirevic:2004ya}
and an earlier result by 
Leutwyler and Roos~\cite{Leutwyler:1984je}.
Our preliminary result is $\Delta f = 0.015(6)(9)$, giving
\bea
{f}_{+}^{K^0\to\pi^-}(0) = 0.962(6)(9),
\label{eq:f0K2pi}
\eea
which agrees well with those of  
Refs.~\cite{Becirevic:2004ya,Leutwyler:1984je}.
Combining with a recent experimental result for 
$|V_{us}|{f}_{+}(0)$~\cite{Alexopoulos:2004sw},
we obtain
\bea
|V_{us}| = 0.2250(14)(20)(12).
\label{eq:Vus}
\eea

\section{Other 4 CKM matrix elements using unitarity and 
Wolfenstein parameters}\label{sec:unitary}

Having the 5 CKM matrix elements directly determined from
the 5 semileptonic decays, we can check unitarity
of the second row of the CKM matrix. Using 
Eqs.~(\ref{eq:VcdVcs}) and (\ref{eq:Vcb}), we get
\bea
({|V_{cd}|^2+|V_{cs}|^2+|V_{cb}|^2})^{1/2}={ 1.00(10)(2)},
\eea
which is consistent with unitarity.
Hereafter the first error is from the lattice calculation and the second
is experimental, unless otherwise stated.

We now use unitarity of the CKM matrix to determine
the other 4 CKM matrix elements.
$|V_{ud}|$, $|V_{tb}|$ and $|V_{ts}|$ are easily determined:
\bea
|V_{ud}| &=& (1 - |V_{us}|^2 - |V_{ub}|^2)^{1/2} ~=~ \Vud, \\
|V_{tb}| &=& (1 - |V_{ub}|^2 - |V_{cb}|^2)^{1/2} ~=~ \Vtb, \\
|V_{ts}| &=& |V_{us}^{*}V_{ub} + V_{cs}^{*}V_{cb}|\ /\ |V_{tb}|
~\simeq~ |V_{cs}^{*}V_{cb}|\ /\ |V_{tb}| ~=~ \Vts. 
\eea
Before proceeding to extract $|V_{td}|$, let us give 
some of the Wolfenstein parameters.
Using Eqs.~(\ref{eq:Vus}), (\ref{eq:Vcb}) and (\ref{eq:Vub}),
we get
\bea
\lambda &=& |V_{us}| ~~~~~~~~~~=\!\!\!\! \Wlambda\!\!\!\!\!\!, \\
A &=& |V_{cb}|/\lambda^2 ~~~~~~=~ \WA, \\
(\rho^2 + \eta^2)^{1/2} &=& |V_{ub}| / (A\lambda^3) ~=~ \WR. 
\label{eq:WR}
\eea

It is difficult to extract $|V_{td}|$ accurately using unitarity 
alone,
because $V_{td}$ ($=A\lambda^3(1\!-\!\rho\!-\!i\eta)$
in Wolfenstein parameterization) contains a large 
CP-violating phase and its magnitude is small.
We use here the experimental result for $\sin(2\beta)$
from $B\to (c\bar{c}) K^{(*)}$ decays. 
By performing a unitary triangle analysis with 
$\sin(2\beta)=0.726(37)$~\cite{HFAG}
and Eq.~(\ref{eq:WR}),
we obtain
\bea
\rho ~=~ \Wrho, &&~ \eta ~=~ \Weta, \\
|V_{td}| &=& \Vtd    
\eea
with a combined lattice and experimental error,
completing the {full} CKM matrix. 

\section{Summary and future plans}\label{sec:summary}

In this paper, we have shown that 
it is indeed possible to fully determine
the CKM matrix using lattice QCD.
$|V_{cd}|$, $|V_{cs}|$, $|V_{ub}|$, $|V_{cb}|$ and $|V_{us}|$
are directly determined
with our lattice results for form factors of
5 semileptonic 
decays.
The error from the quenched approximation is removed by
using the MILC unquenched gauge configurations.
The error from the $m_l\to m_{ud}$ extrapolation 
is significantly reduced by using improved staggered quarks.
The accuracy is comparable to 
that of the Particle Data Group averages, as seen in
Table~\ref{tab:error}.
The other 4 CKM matrix elements
$|V_{ud}|$, $|V_{tb}|$, $|V_{ts}|$ and $|V_{td}|$ are then determined
by using unitarity of the CKM matrix and the experimental result
for $\sin{(2\beta)}$.
In this way, we obtain the full CKM matrix, Eq.~(\ref{ckm}),
and the Wolfenstein parameters, Eq.~(\ref{Wolf}).
We emphasize that the only theoretical input in this paper
is lattice QCD. 
This concretely shows that lattice QCD is one of the most powerful tools
in aiding our understanding of flavor physics.

As for future works, we are planning to improve the accuracy of 
form factors of $D\to \pi(K) l\nu$ and $B\to \pi l\nu$ decays,
which is currently $\O(10\%)$.
To reduce the largest systematic error from the lattice discretization
effects, we are repeating the calculations at a smaller lattice 
spacing.
The unquenched calculations of leptonic decays 
($D_{(s)}\to l\nu$, $B_{(s)}\to l\nu$) are 
also underway~\cite{Simone:2004fr,Wingate:2003gm}.

To test the Standard Model,
we need to calculate 
mixing parameters of neutral $B$ and $K$ mesons, $B_B$ and $B_K$.
The unquenched ($n_f=2+1$) calculation of $B_B$ and $B_K$ 
is underway by the HPQCD collaboration~\cite{Gamiz:2004qx},
and 
is planned by the Fermilab Lattice collaboration 
and by the LANL lattice group.
The initial accuracy will be $\O(10\%)$, and we hope to have 
an accuracy of 5\% in the near future.
We will then be able to have tighter constraints on the
$(\rho,\eta)$ plane, leading to a precision test of the Standard Model,
and guiding the search for new physics.

\section*{Acknowledgments}

We thank all members of the Fermilab Lattice, MILC and HPQCD Collaborations,
in particular, the authors in Ref.~\cite{Aubin:2004ej};
C.~Aubin,
C.~Bernard,
C.~DeTar,
M.~Di Pierro,
A.~El-Khadra,
Steven~Gottlieb,
E.~B.~Gregory,
U.~M.~Heller,
J.~Hetrick,
A.~S.~Kronfeld,
P.~B.~Mackenzie,
D.~Menscher,
M.~Nobes,
M.~B.~Oktay,
J.~Osborn,
J.~Simone,
R.~Sugar,
D.~Toussaint, and 
H.~D.~Trottier.
We thank Andreas~S.~Kronfeld, Jack~Laiho, and Robert~Sugar
for valuable comments on the manuscript.
This work is supported by
the Fermilab Computing Division, the SciDAC program, 
the Theoretical
High Energy Physics Programs at the DOE and NSF, and URA.

%
\label{MOkamotoEnd}

\end{document}